\documentstyle[twocolumn,aps,prl]{revtex}

\input psfig

\begin{document}

\wideabs{
\draft
\title{Different Hagedorn temperatures
for mesons and baryons from experimental
mass spectra, compound hadrons, and combinatorial saturation}
\author{Wojciech Broniowski and Wojciech Florkowski}
\address{H. Niewodnicza\'nski Institute of Nuclear Physics, 
PL-31342 Cracow, Poland}
\maketitle
\begin{abstract}
We analyze the light-flavor particle mass spectra
and show that in the region up to $\simeq 1.8$GeV
the Hagedorn temperature for baryons is about 30\% smaller
than for mesons, reflecting the fact that the number of baryon
states grows more rapidly with the mass. We also show that the spectra
are well reproduced in a model where hadrons are
compound objects of quanta, whose available number increases with mass.
The rapid growth of number of hadronic states is a combinatorial effect.
We also point out that an upper limit on the excitation energy
of these quanta results in a maximum number of hadron states that can be
formed. According to this combinatorial saturation,
no more light-flavor
hadron resonances exist above a certain mass.
\end{abstract}
}

In 1965 Hagedorn \cite{hagedorn} postulated that for large masses $m$ the
spectrum of hadrons grows exponentially, $\rho (m)\sim \exp \left(
m/T_{H} \right) $, where $T_{H}$, the Hagedorn temperature, is a scale
parameter. The hypothesis was based on the observation that at some point a
further increase of energy in $pp$ and $p\bar{p}$ collisions no longer
raises the temperature of the formed fireball, but results in more and more
particles being produced. Thus, there is a maximum temperature that a
hadronic system can achieve. The Statistical Bootstrap Model \cite{SBM1,SBM2}
predicted that asymptotic behavior, namely $\rho (m)\simeq cm^{a}\exp \left(
m/T_{H}\right) $, where $a$ is a negative power ($a\leq -5/2$ \cite
{SBM1}, or $a\leq -3$ \cite{SBM2}). The parameters of the asymptotic
spectrum are to be determined by comparing to the experimental mass spectra.
The fits of the sixties had a rather poor spectrum to their disposal,
sufficiently dense only in the range of masses up to $1$GeV, and sparse
above. Still, Hagedorn and Ranft \cite{HagRan} were able to fit the function
\begin{equation}
\rho _{HR}(m)=\frac{c}{\left( m^{2}+m_0^{2}\right) ^{5/4}}\exp
\left( m/T_{H}\right) ,  \label{rhoHR}
\end{equation}
with $m_0=0.5$GeV,
and got $T_{H}=160$MeV. Frautschi \cite{Frautschi} obtained similar values for
the Hagedorn temperature. Ever since it was believed that there is one
universal scale in the asymptotic spectrum of hadrons, and its value is
about $160$MeV. Meanwhile, the particle tables became more complete and
longer, with a total of 3182 light-flavor states (counting mesons, baryons,
and antibaryons with their spin and isospin degeneracies) compared to 1432
used in Ref. \cite{HagRan}. This allows for a much more stringent
verification of the Hagedorn hypothesis. We have done this, with surprising
results.

We have used the 1998 edition of the Particle Data Group tables \cite{PDG}.
Rather then comparing the experimental and theoretical spectra, we compare
their {\em cumulants}: the number of states with the energy less than $m$, 
\begin{equation}
N_{{\rm exp}}(m)=\sum_{i}g_{i}\Theta (m-m_{i}),  \label{Nm}
\end{equation}
and 
\begin{equation}
N_{{\rm theor}}(m)=\int_{0}^{m}\rho _{{\rm theor}}\left( m^{\prime }\right)
dm^{\prime }  \label{Nth}
\end{equation}
where $\Theta $ is the step function, $m_{i}$ is the mass of the resonance,
and $g_{i}$ is its degeneracy. Obviously, $dN(m)/dm=\rho \left( m\right) $.
Although the information content is the same whether one uses $\rho (m)$
directly, or $N(m)$, the use of cumulants is advantageous, since that way
one avoids the nuisance of producing histograms of $\rho _{{\rm exp}}$. Our
results are presented in Fig. 1, where we show the $\log _{10}N(m)$
separately for mesons and baryons. We can clearly sea that up to $m\simeq 1.8
$GeV the data line-up along (almost) straight lines.
The solid lines are  the least-square
fits to the data according to Eq. (\ref{rhoHR}) over the range up to $1.8$%
GeV, skipping the lightest particle in the set.
The behavior of Fig.1 is compatible with
the Hagedorn hypothesis.

\begin{figure}[tbp]
\centerline{%
\psfig{figure=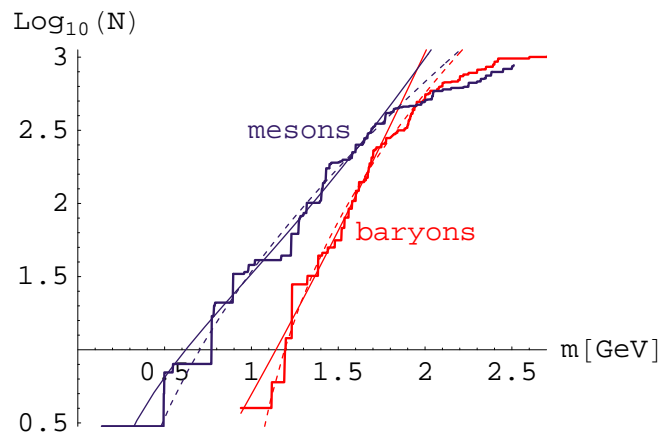,height=5.7cm,%
bbllx=82bp,bblly=384bp,bburx=531bp,bbury=680bp,clip=}}
\vspace{2mm}
\caption{Cumulants of meson and baryon spectra, Hagedorn fits (solid lines),
and Compound Hadron Model fits (dashed lines), plotted as functions of mass.}
\end{figure}

However, the slopes in Fig. 1 are different for mesons and baryons, which
means {\em different Hagedorn temperatures for mesons and baryons} for
masses in the region up to $1.8$GeV. This is the key observation of this
paper. Applying the form (\ref{rhoHR}), with $m_{0}=500$MeV, we find the best
fit for the mesons with $T_{{\rm mes}}=197$MeV, and for the baryons with $T_{%
{\rm bar}}=141$MeV. These temperatures differ considerably from
each other, reflecting the
fact that the {\em baryon spectrum grows much more rapidly}.

A question arises if it were possible that meson states are missed in
experiments much more frequently than baryons, such that the effect of Fig.
1 is spurious. The numbers show this is highly unlikely. In order to make
the meson line parallel to the baryon line we would have to aggregate as
much as $500$ additional states up to $m=1.8$GeV, more than the present
number of $407$ states at this point. Thus, we reject the scenario of
missing meson states. We also remark that discoveries of some extra states
can only increase the slopes in Fig. 1, thus lowering the presently-fitted
values of $T_{{\rm mes}}$ and $T_{{\rm bar}}$. Another question is whether
we have reached the asymptotics with $m\sim 1.8$GeV. This question can only
be addressed within a sufficiently accurate theory or model.
It should be stressed that the function multiplying the exponent
has significance for the fitted values of
$T_{{\rm mes}}$ and $T_{{\rm bar}}$. For instance, with $m_0=1$GeV
we obtain 228MeV and 152MeV
for mesons and baryons, respectively. Using different values of $m_0$
for mesons and baryons may bring the two temperatures closer to each other.
Anyway, in the
range of $m$ up to $\sim 1.8$GeV we {\em effectively}
have two distinct Hagedorn
temperatures for mesons and baryons, as seen in Fig. 1.

\begin{figure}[tbp]
\centerline{%
\psfig{figure=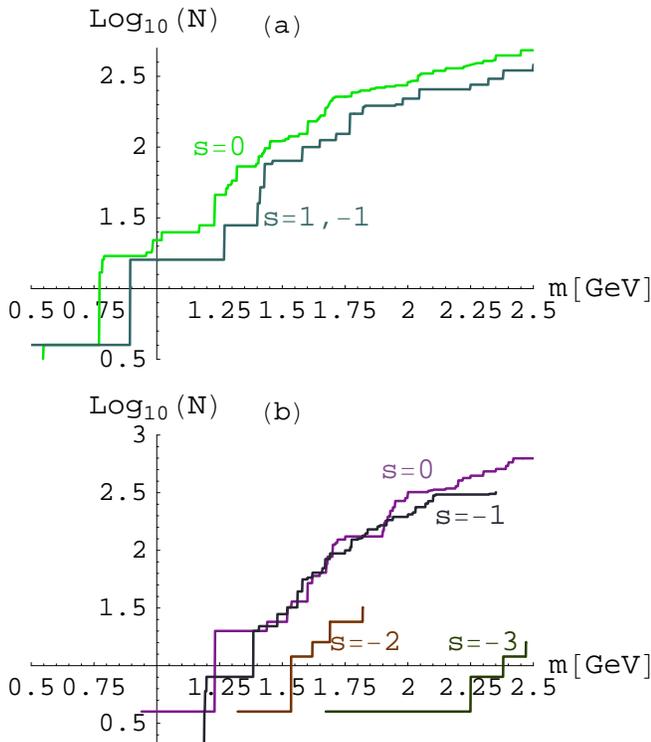,height=9.9cm,%
bbllx=92bp,bblly=200bp,bburx=520bp,bbury=690bp,clip=}}
\caption{Cumulants of meson (a) and baryon (b) spectra with selected
strangeness, plotted as functions of mass.}
\end{figure}

In Fig. 2 we compare the cumulants of spectra for mesons and baryons of
definite strangeness. We can see the {\em universality of slopes} for both
mesons and baryons.

The theoretical challenge is to explain the different behavior of mesonic
and baryonic spectra. Historically, the Dual String Model \cite{Jacob}
produced the exponentially-growing mass spectra. In the framework of the
Veneziano model \cite{Veneziano} Huan and Weinberg \cite{HuanW} derived
the following asymptotic formula:
\begin{eqnarray}
\rho _{V}(m) &\sim &2\alpha ^{\prime }mP_{D}(n),  \label{rhoHW} \\
P_{D}(n) &=&\sqrt{\frac{1}{2n}}\left( \frac{D}{24n}\right) ^{\frac{D+1}{4}%
}\exp \left( 2\pi \sqrt{\frac{Dn}{6}}\right) ,  \label{PD} \\
n &=&\alpha _{0}+\alpha ^{\prime }m^{2},  \label{rhoHW2}
\end{eqnarray}
where $\alpha ^{\prime }$ is the slope of the Regge trajectories, $\alpha
_{0}$ is the intercept (averaged over various trajectories), and $D$ is the
number of dimensions, usually assumed to be $4$ \cite{HuanW}, although
higher values have also been attempted \cite{Lovelace}. In the strict
sense, the dual string models are constructed for mesons, and their
application
to baryons has a more phenomenological character.
Since $\alpha^{\prime }\approx 1$GeV$^{-2}$ is universal for all hadrons, the difference
between mesons and baryons could only come from different values of $\alpha
_{0}$, and possibly from $D$. We have tried all reasonable values for these
parameters, and concluded that they result in very similar curves.
The function multiplying the exponent in Eq. (\ref{rhoHW}) has
significance here. For larger $D$ it decreases faster with $m$,
which effectively compensates for the growth from the exponent in
the interesting region of $m$. We were able to fit
the meson line in Fig. 1 with formula (\ref{rhoHW}),
but it is impossible to fit the baryons.
Moreover, the
experimentally accessible range of $m$ is not asymptotic enough to justify
Eq. (\ref{rhoHW}). Indeed, Eq. (\ref{rhoHW}) is a direct consequence of a
combinatorial problem known as {\em partitio numerorum}. For $D=1$ Eq. (\ref
{rhoHW}) is the asymptotic form for the number of different partitions of a
positive integer $n$ into non-negative integer components \cite{partitio}.
For $D>1$ it is a generalization to the case where the components are of $D$
different types \cite{HuanW}. From the Regge formula (\ref{rhoHW2}) we can
see that for $m$ in the range $1-2$GeV the values of $n$ lie between $1$ and
$4$, hence $n$ is not large enough to justify the asymptotic form
(\ref{rhoHW}). In
this paper we do not pursue the dual-model track any further, and leave it
with the conclusion
that {\em it works for the meson spectrum}, and fails, in the present form,
for baryons.

Is there a simple explanation of the behavior of Fig. 1?
What we have to search for is a mechanism of the observed rapid
increase of degrees of freedom with $m$. Inspired by the combinatorics of 
{\em partitio numerorum} we propose a candidate mechanism capable of
explaining the result of Fig. 1 not worse than the Hagedorn model. We call
it the {\em Compound Hadron Model}, in analogy to compound nuclei
\cite{compound}. In the statistical model of nuclear reactions a compound nucleus
is an object whose density of states is enormous, and is described
approximately by the formula $\rho (E^{\ast })\simeq f(E^{\ast })\exp \left(
b\sqrt{E^{\ast }}\right) $, where $E^{\ast }$ is the (large) excitation
energy of the nucleus, $b$ is a nucleus-dependent constant, and $f(E^{\ast })
$ is a slowly-varying function. Let us illustrate how this formula can arise
in a toy example. Suppose for simplicity that we can neglect the exclusion
principle, and the system has equally-spaced single-particle excitation
levels, $\varepsilon _{k}=k\Delta E$. In the ground state all particles sit
in the lowest level, $\varepsilon _{0}=0$. The excitation energy can be
composed in very many different ways. Let us express $E^{\ast }$ in units of 
$\Delta E$, i.e. $E^{\ast }=n\,\Delta E$. We can make $E^{\ast }$ by
exciting $n$ times the lowest (first) level, $E^{\ast }=n\varepsilon _{1}$,
once the second level and $n-2$ times the first level, $E^{\ast
}=\varepsilon _{2}+(n-2)\varepsilon _{1}$, and so on. In general, $E^{\ast
}=\varepsilon _{1}k_{1}+\varepsilon _{2}k_{2}+...+\varepsilon _{n}k_{n}$,
which means $n=k_{1}+2k_{2}+...+nk_{n}$, with integer $k_{i}\geq 0$. Of
course, this is nothing else but the partition problem, and the number of
distinct ways this may be achieved is, for a large $n$, given by Eq. (\ref{PD}%
) with $D=1$. The difference is, however, that now $n$ is proportional to
the excitation energy, and not the square of mass, as in Eq. (\ref{rhoHW2}).
One can extend this combinatorial problem to account for $D$ different types
of excitations: $n=\sum_{\alpha }k_{1}^{(\alpha )}+2\sum_{\alpha
}k_{2}^{(\alpha )}+...+n\sum_{\alpha }k_{n}^{(\alpha )}$ , $\alpha =1,..,D$, 
$k_{i}^{(\alpha )}\geq 0$, which gives the number of partitions, $P_{D}(n)$,
of Eq. (\ref{PD}). In addition, we may incorporate the exclusion principle
by disallowing the repetition of the same level, {\em i.e.} requesting that
none of the numbers $k_{i}^{(\alpha )}$ is repeated. This leads to the
asymptotic formula \cite{Narkiewicz} 
\begin{equation}
P_{D}^{{\rm NR}}(n)\sim n^{-3/4}\exp \left( 2\pi \sqrt{\frac{D}{12}n}\right)
,  \label{pnnr}
\end{equation}
where ``NR'' indicates ``no repetitions''. Note the factor of $\frac{1}{12}$
under the square root compared to $\frac{1}{6}$ in Eq. (\ref{PD}) (there are
less partitions without repetitions than with repetitions), and the absence
of $D$ in the power $n$ multiplying the exponent.

Inspired by the compound nucleus model and by the expression for $P_{D}^{%
{\rm NR}}(n)$, we propose the following guess formula for the hadronic mass
spectra, which is a direct consequence of Eq. (\ref{pnnr}):
\begin{equation}
\rho (m)=\frac{c\Theta (m-m_{0})\exp \left( 2\pi \sqrt{\frac{D}{12}\frac{%
\left( m-m_{0}\right) }{\Delta E}}\right) }{\left( \left( m-m_{0}\right)
^{2}+(0.5{\rm GeV})^{2}\right) ^{5/8}},  \label{chmrho}
\end{equation}
where $c$ is a constant, $m_{0}$ is the ground-state mass, $D$ is the number
of types of excitations, and $\Delta E$ is the average level spacing.
The asymptotic power of the function multiplying the exponent is $-5/4$,
whereas the power $-2$ is used in an analogous formula in compound
nuclei \cite{compound}. The underlying physical picture is
following: hadrons are bound objects of constituents (quarks, gluons,
pions).
The Fock space contains a ground state, and excitations on top of it. In the
case of the compound nucleus these elementary excitations
are $1p1h$, $2p2h$%
, $3p3h,$ {\em etc.} states. In the case of hadrons they are formed of $q%
\bar{q}$ and gluon excitations, {\em e.g.} for mesons we have $q\bar{q}$, $q%
\bar{q}g$, $qq\bar{q}\bar{q}$, $q\bar{q}gg$, {\em etc.} Now, we form the
excitation energy (hadron mass) by differently composing elementary
excitations, which bring us to the above-described partition problem. It
seems reasonable to take zero ground-state energy for mesons, $m_{0}^{{\rm %
mes}}=0$, which are excitations on top of the vacuum. For baryons we take $%
m_{0}^{{\rm bar}}=900$MeV, which is basically the mass of the nucleon. The
quantity $\Delta E/D$ is treated as a model parameter and is fitted to data.

The results of the compound-hadron-model fit, Eq. (\ref{chmrho}), are shown
with the dashed line in Fig. 1. The curves are slightly bent down, compared
to the Hagedorn-like fits (solid lines), which is caused by the square root
in the exponent of Eq. (\ref{chmrho}). But the fits are no worse, or even
better when the fit region is extended to $m=2$GeV. Numerically, the
least-square fit for $m$ up to 1.8GeV
gives $\Delta E^{{\rm mes}}/D=50$MeV for
mesons, and $\Delta E^{{\rm bar}}/D=53$MeV for baryons.
The proximity of these numbers shows that the scales for
mesons and baryons are similar.
The number of types of the excitations, $D$,
is taken to be 1. This means that we treat all levels, carrying
spin, orbital, radial, or isospin quantum numbers, as separate.
If we treat them as D-times degenerate, then the average spacing
$\Delta E$ increases $D$ times, such that the dependence on
$D$ cancels. Note that in case of gluonic or $q\bar{q}$ excitations
color does not bring extra degeneracy, since all states have
to be color-neutral and there is only one way in which the
elementary excitations can be
coupled to a color singlet.
The obtained values for $\Delta E^{{\rm mes}}$ (with $D=1$) mean that
the corresponding $n$ at $m=1.8$GeV is around $35$ for mesons
and $17$ for baryons. Such $n$ are
sufficiently large to justify the use of the
asymptotic formula (\ref{pnnr}).

There is an interesting effect we wish to point out. It is natural to expect
that a bound hadronic system has an upper limit for the excitation energy.
It is helpful to think here of bags with a finite depth, or of breaking-up
flux-tubes. Thus, in constructing the Fock space for bound objects we should
have a limited number of quanta to our disposal. Translating this idea into
our toy combinatorial problem, we now have to ask about the number of ways
of partitioning $n$ (without repetitions) into integer components, $%
n=\sum_{\alpha }k_{1}^{(\alpha )}+2\sum_{\alpha }k_{2}^{(\alpha
)}+...+n\sum_{\alpha }k_{n}^{(\alpha )}$ , but now with $k_{{\rm max}}\geq
k_{i}^{(\alpha )}\geq 0$, where $k_{\rm max}$
is the maximum available number. We
denote this number of partitions with limited $k_{i}^{(\alpha )}$ as $P_{D}^{%
{\rm NR}}(n,k_{{\rm max}})$. Figure 3 shows the decimal $\log $ of cumulants
of this quantity 
for $D=1$, $\sum_{i=1}^{n}P_{1}^{{\rm NR}}(i,k_{{\rm max}})$, plotted
as a function of $n$ for various values of $k_{{\rm max}}$. The case $k_{%
{\rm max}}=\infty $ corresponds to partitions with unlimited $k_{i}^{(\alpha
)}$. Each curve becomes flat at $n=k_{{\rm max}}(k_{{\rm max}}+1)/2$, which
is the point where all available levels have been filled. Beyond that point
we cannot form bound states any more. We term this effect {\em combinatorial
saturation}. It means that in a compound-hadron model beyond a certain
mass $m$ no more light-quark resonances can be formed. For instance,
our toy model with $D=1$ and $k_{{\rm max}}=12$
predicts that this happens at $n=78$.
With $\Delta E$ obtained in the fits, this leads,
according to the formula $m=m_{0}+n\Delta E$,
to a maximum mass of $\sim 4$GeV for light-quark mesons and $\sim 5$GeV for
light-quark baryons. We can see in Fig. 3 that before reaching the
combinatorial saturation, the curves with a given $k_{{\rm max}}$ depart
from the unlimited-partition curve, $k_{{\rm max}}=\infty $. It turns out
that the $k_{{\rm max}}=\infty $ curve is very close to the asymptotic
(large $n$) curve, described by formula (\ref{pnnr}), even for low values of
$n$. This is why Eq. (\ref{chmrho}) can be used even at low values of $m$.
We may thus speculate that the departure of data from the dashed lines in
Fig. 1 around $m=2$GeV, similar to the behavior of Fig. 3, is a signal of
the combinatorial saturation. More complete data
in that region and above would definitely help to check 
if this is really the case.
We note that Bisudov\'a, Burakovsky and Goldman \cite{bbg}
obtain similar limits
for mass of light hadrons from non-linear Regge trajectories.

\begin{figure}[tbp]
\centerline{%
\psfig{figure=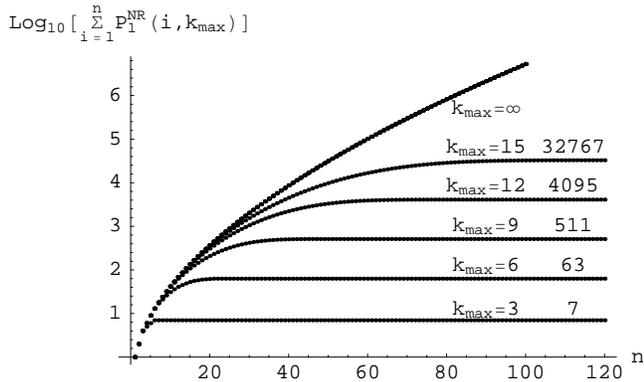,height=5.1cm,%
bbllx=78bp,bblly=426bp,bburx=558bp,bbury=713bp,clip=}}
\vspace{2mm}
\caption{Cumulants of $P_{1}^{{\rm NR}}(i,k_{{\rm max}})$ for
various values of $k_{{\rm max}}$. The integers at the
right ends of the curves are the saturation values of the cumulants.}
\end{figure}

Let us summarize the basics of combinatorics
behind the formation of the hadronic spectra. 
As we increase the mass, more elementary 
excitations may participate in the resonance-formation 
process, in other words 
{\em more degrees of freedom} open up. This,  
via partitio numerorum, leads
to an exponential growth of the number possibilities of forming 
excited states. Beyond a certain mass there are no more elementary 
excitations available, and we reach the combinatorial saturation in the number
of hadronic states. Clearly, the real life of hadronic physics is more 
complicated than the toy model discussed above. However, the 
result may be robust and not care about the details, for example whether
the levels are equally spaced, as assumed here, or not.
The success of the statistical model in nuclear
physics shows that details are not needed to understand gross features
system with many degrees of freedom.

Finally, let us comment on the different
power of $m$ in the exponent of Eq. (\ref{chmrho}) and the Hagedorn hypothesis,
Eq. (\ref{rhoHR}). It is not a paradox, since our simple
assumptions in the language of quark and gluon (or pion) excitations,
are completely different than the assumptions of the Statistical
Bootstrap Model, where self-similarity of fireballs is the key
ingredient.
The thermodynamical implications of the two models are different:
whereas in the Hagedorn model the temperature $T$ of
a hadronic system cannot be larger than the Hagedorn
temperature, in the Compound Hadron Model
there is no formal limit on $T$.
We stress that this is of formal significance only,
since deconfinement is expected to occur at
temperatures of the order of $\sim 150$MeV, dissolving hadron resonances
into the quark-gluon plasma.

One of us (WB) is grateful to A. Bia\l{}as, A. Horzela, J. Kwieci\'nski,
and P. \.Zenczykowski for useful discussions.


\end{document}